\documentclass[twocolumn,secnumarabic, amssymb,
floatfix,amsmath,tightenlines,showpacs,superscriptaddress,
preprintnumbers,aps,prl]{revtex4-1}

\usepackage{times}
\usepackage{color}
\usepackage{graphicx}% Include figure files
\usepackage{dcolumn}% Align table columns on decimal point
\usepackage{amssymb}
\usepackage{amsmath}
\usepackage{amsfonts}
\usepackage{docs}%
\usepackage{bm}%
\usepackage[leftcaption]{sidecap}
\usepackage{wrapfig}
\usepackage[colorlinks=true,linkcolor=blue,pagecolor=blue,filecolor=blue,menucolor=blue,urlcolor=blue,citecolor=blue,anchorcolor=blue]{hyperref}%
\newcommand{\etal}{\emph{et al.}}

\newcommand{\ff}{$F$/$F$ }
\newcommand{\fs}{$S$/$F$ }

\begin{document}

\title{
Long-Range Triplet Supercurrents Induced by Singlet Supercurrents
Parallel to Magnetic Interfaces}

\author{Mohammad Alidoust }
\email{phymalidoust@gmail.com}
\affiliation{Department of Physics,
University of Basel, Klingelbergstrasse 82, CH-4056 Basel, Switzerland}
\affiliation{Department of Physics,
Faculty of Sciences, University of Isfahan, Hezar Jerib Avenue,
Isfahan 81746-73441, Iran}
\author{Klaus Halterman}
\email{klaus.halterman@navy.mil}
\affiliation{Michelson Lab, Physics
Division, Naval Air Warfare Center, China Lake, California 93555,
USA}

\date{\today}

\begin{abstract}
Employing a spin-parameterized Kleldysh-Usadel technique for the
diffusive regime, we demonstrate that even in the low proximity
limit, considerable long-ranged triplet supercurrents can be
effectively generated by spin-singlet supercurrents flowing
\textit{parallel} to
the interfaces of uniform
double ferromagnet interlayers
with noncollinear exchange
fields ``{\it independent}" of actual junction geometry. The triplet
supercurrents are found to be most pronounced when the thicknesses
of the ferromagnet strips are unequal.
To experimentally verify this generic phenomenon,
we propose an accessible and well-controlled structure that can
fully isolate the long-range triplet effects.

\end{abstract}


\maketitle

Spin carriers play a crucial role in spintronics devices where the
spin of the carriers are used to transport or store information
\cite{zutic}. The spin current in diffusive ferromagnets due to
spin-polarized charged carriers has a short-range propagation and
limited spin coherence, thus presenting a detriment to the
functionality of most spintronics devices \cite{Kimura_prl}. This
obstacle, however, can be resolved within the cryogenic realm by
utilizing ferromagnetic ($F$)/superconductor ($S$) hybrids to generate
spin polarized superconducting correlations that can propagate over
several hundred nanometers with limited decay
\cite{golubov_rmp,buzdin_rmp,Quay,bergeret_rmp}.
The induced spin-triplet odd-frequency superconducting correlations
were theoretically predicted \cite{bergeret_rmp} for systems
comprised of an $s$-wave superconductor in proximity to a
ferromagnet. The signatures of such long-range proximity
superconducting spin correlations were found shortly
thereafter in experiments %khf
\cite{ryaz0}. Numerous works since then, both experimentally and
theoretically, have been devoted to the study and generation of
superconducting correlations with net spin
\cite{norm,robinson,houzet,paral_sffs,paral_neel}.

It has been shown that it is possible to study long-range
supercurrents in \textit{ballistic} bilayer \cite{Trifunovic} and
\textit{diffusive} trilayer \cite{houzet} magnetic Josephson
junctions where the supercurrent flows transversely relative to
the noncollinear $F/F$ interfaces.
The situation is unfavorable however in the \textit{low proximity}
limit \cite{alidoust_sffn} where only faint signatures of the
triplet supercurrents with spin projection $m$=$\pm 1$ on the
quantization axis exist, even in uniformly magnetized trilayers
\cite{houzet,alidoust_sffn}. The low proximity limit is realized for
low transparency \fs interfaces, strongly magnetized layers, thick
$F$ barriers, and temperatures near the superconducting critical
temperature, which prevails in many experiments
\cite{bergeret_rmp,buzdin_rmp}.
Thus, it is of considerable interest to determine the
most simple and generic
situations that permit generation of long-range
triplet supercurrents in the low proximity limit. Moreover, the
findings in this limit may pave the way for more pronounced triplet
generation in similar structures within
the ballistic and full proximity limit of diffusive regime
\cite{bergeret_rmp,buzdin_rmp}.

\begin{figure*}[]
\includegraphics[width=17.30cm,height=2.8cm]{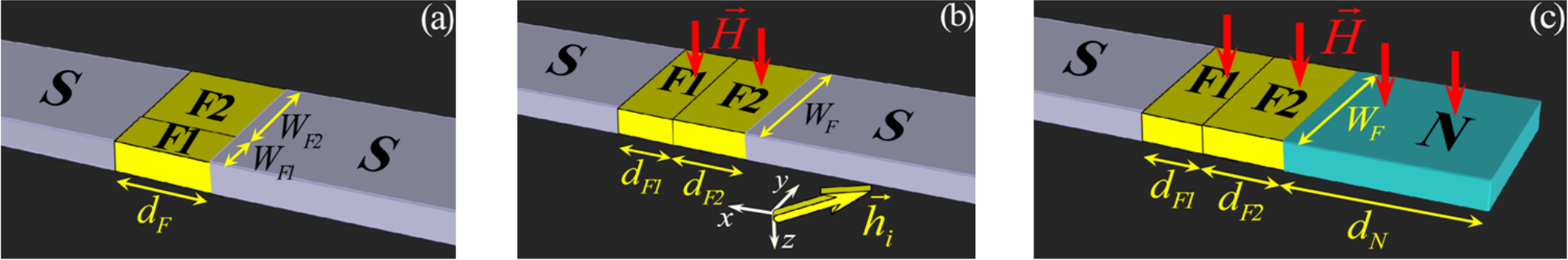}
\caption{\label{fig:model1} Schematic of the proposed experimental
setups of the basic $F/S$ hybrid structures supporting supercurrent
parallel to the $F_1/F_2$ interfaces. The ferromagnetic layers,
$F_1$ and $F_2$ have uniform magnetizations and $N$ represents a
nonmagnetic normal metal with a finite thickness $d_N$.
Superconductivity is induced via $s$-wave superconducting terminals
of infinite extent (labeled $S$). We denote the exchange field of
each $F$ layer by
$\vec{h}_i$$=$$(h_i^x,h_i^y,h_i^z)$$=$$h_0(\sin\beta_i\cos\alpha_i,\sin\beta_i\sin\alpha_i,\cos\beta_i)$,
in which $\beta_i$ and $\alpha_i$ are spherical angles for
$i$$=$$1,2$. The junctions are located in the $xy$ plane and the
$S$/$F$ interfaces are along the $y$-axis. In panel (a) the
interface of the double ferromagnet region is oriented perpendicular
to the $S$/$F$ interfaces creating a two-dimensional hybrid. The
widths of the rectangular $F$ strips $W_{F1}$, and $W_{F2}$ are
generally different, while their lengths, $d_F$, are equal. In (b)
and (c) all interfaces are parallel to the $y$ axis. The widths of
the rectangular $F$ strips, $W_F$, are the same while the length of
each magnetic layer, $d_{F1}$ and $d_{F2}$, are not necessarily
equal. To generate a supercurrent flowing along the $F_1/F_2$
interface, an external magnetic field $\vec{H}$ is directed along
the $z$-axis, normal to the junction plane.}
\end{figure*}

In this Letter, we demonstrate that by considering singlet
supercurrent flow ``parallel'' rather than ``transverse'' to the
$F/F$ interfaces, substantial long-range triplet supercurrents can
be induced
{\it regardless} of actual configuration and geometry
\cite{paral_neel,paral_sffs}. We make use of the two-dimensional
quasiclassical Keldysh-Usadel approach in configuration-space, which
is a general formalism appropriate for inhomogenous diffusive $F/S$
heterostructures containing generic magnetization and external
magnetic field profiles \cite{alidoust_nfrh,alidoust_nfrh1}. This two-dimensional
approach is capable of explorations previously unattainable in
strictly one-dimensional systems. A spin-parametrization technique
is also incorporated within our method to pinpoint the behavior of
the odd- and even-frequency components of the total supercurrent. To
demonstrate the {\it generality} of our findings,
we consider a variety of structures \cite{paral_sffs,paral_neel} and
specifically establish clear insights into the spatial profiles
of singlet/triplet supercurrents in three different magnetic
structures that host parallel supercurrent flow: A noncollinear
double $F_1/F_2$ interlayer that is either vertically
[Fig.~\ref{fig:model1}(a)] or horizontally
[Fig.~\ref{fig:model1}(b)] stacked between two $S$ electrodes, or
one that is horizontally sandwiched between a superconducting electrode and a
finite sized normal metal [Fig.~\ref{fig:model1}(c)]. In order to
have supercurrent flow parallel to the $F_1/F_2$ interfaces in the
two latter structures, we introduce an externally applied magnetic
field normal to the junction plane. The structures reveal that
short-range spin-singlet supercurrents flowing parallel to the \ff
interfaces containing noncollinear magnetizations and unequal
thicknesses are effectively converted into long-ranged spin-triplet
supercurrents.
The simple structure shown in
Fig.~\ref{fig:model1}(c),
permits experimental isolation of the long-range triplets
and confirmation of this generic effect.

A general three-dimensional quasiclassical formalism for diffusive
magnetic heterostructures subject to an external magnetic field is
described by the following Usadel equations
\cite{cite:usadel,alidoust_nfrh,alidoust_nfrh1}:
\begin{align}\label{eq:full_Usadel}
D[\check{\partial},\check{G}[\check{\partial},\check{G}]]+i[
\varepsilon +
\text{diag}[\vec{h}\cdot\vec{\tau},(\vec{h}\cdot\vec{\tau})^{T}],\check{G}]=0,
\end{align}
in which the Pauli matrices constitute the components of
$\vec{\tau}$. We denote the diffusion constant of the medium by $D$
and $\check{G}$, $\vec{h}$ are functions of coordinates ${\bm
r}$$\equiv$$(x,y,z)$ \cite{alidoust_nfrh,alidoust_nfrh1}. Here, the exchange field
describing the ferromagnetic region, $\vec{h}({\bm r})$$=$$(h^x({\bm
r}),h^y({\bm r}),h^z({\bm r}))$, can take arbitrary directions. We
have defined a 4$\times$4 partial derivative matrix,
$\check{\partial}$, as $\hat{\partial}$$\equiv$$\vec{\nabla}
\hat{1}$$-$$ie \vec{A}({\bm r})\hat{\rho}_{3}$, in which $\vec{A}$
is the vector potential of the applied magnetic field, $\vec{H}$.
The energy, $\varepsilon$,  of the quasiparticles  is measured from
the Fermi surface $\varepsilon_F$. The resultant sixteen coupled
complex partial differential equations should be supplemented by the
appropriate boundary conditions to properly capture the electronic
and transport characteristics of $F/S$ hybrid structures. We employ
the Kupriyanov-Lukichev boundary conditions at the \fs interfaces
\cite{cite:zaitsev} and the induced proximity correlations are tuned
by the barrier resistance $\zeta$: $
\zeta(\check{G}\check{\partial}\check{G})\cdot\hat{\boldsymbol{n}}$$=$$[\check{G}_{\text{BCS}},\check{G}]$,
in which $\hat{\boldsymbol{n}}$ is a unit vector normal to the
interfaces and $\check{G}_{\text{BCS}}$ is the superconducting bulk
solution. Under equilibrium conditions, the vector current density
is expressed as an integration of the Keldysh block, $\vec{J}({\bm
r})$$ =$$ J_{0} \int d\varepsilon\text{Tr}\bigl\{\rho_{3}
\big(\check{G} [\check{\partial},\check{G}]\bigr)^{K}\}$, where
$J_{0} $$ =$$ N_{0} e D/4$, $N_{0}$ is the density of states at the
Fermi surface, and $e$ is the electron charge. To gain insight into
the triplet supercurrent components, we use the spin-parametrization
scheme for the Green's function \cite{bergeret_rmp}. Therefore, the
anomalous component of the Green's function takes the following form
in terms of the even  ($\mathbb{S}$) and odd  ($\mathbb{T}$)
frequency parts: $F({\bm r},\varepsilon)$$=$$ i\big [\mathbb{S}({\bm
r},\varepsilon)$$+$$\vec{\mathbb{T}}({\bm
r},\varepsilon).\vec{\tau}\big ]\tau_y$, where
$\vec{\mathbb{T}}({\bm r},\varepsilon) \equiv
(\mathbb{T}_x,\mathbb{T}_y,\mathbb{T}_z )$, and
$\mathbb{T}_x,\mathbb{T}_y$ have $m$=$\pm 1$ projections along
the quantization axis, while $\mathbb{T}_z$ has
$m$=$0$ \cite{bergeret_rmp}.
If we now substitute this
decomposition into the Usadel equation, Eq.~(\ref{eq:full_Usadel}),
we are able to separate the contributions $\mathbb{S},
\mathbb{T}_x,\mathbb{T}_y$, and $\mathbb{T}_z$ to the charge
supercurrent. To study the precise behavior of individual components
constituting the Josephson current, we introduce the following
decomposition scheme for the current density:
$J_{x,y}$$=$$J^{s0}_{x,y}$$+$$J^{sx}_{x,y}$$+$$J^{sy}_{x,y}$$+$$J^{sz}_{x,y}$,
where $J^{s0}_{x,y}$ encompasses  the $\mathbb{S}$ terms, and $J^{s
\gamma}_{x,y}$  the $\mathbb{T}_\gamma$ terms (designating
$\gamma$$=$$x,y,z$).
\begin{SCfigure*}[1][h]
\includegraphics[width=12.5cm,height=5.0cm]{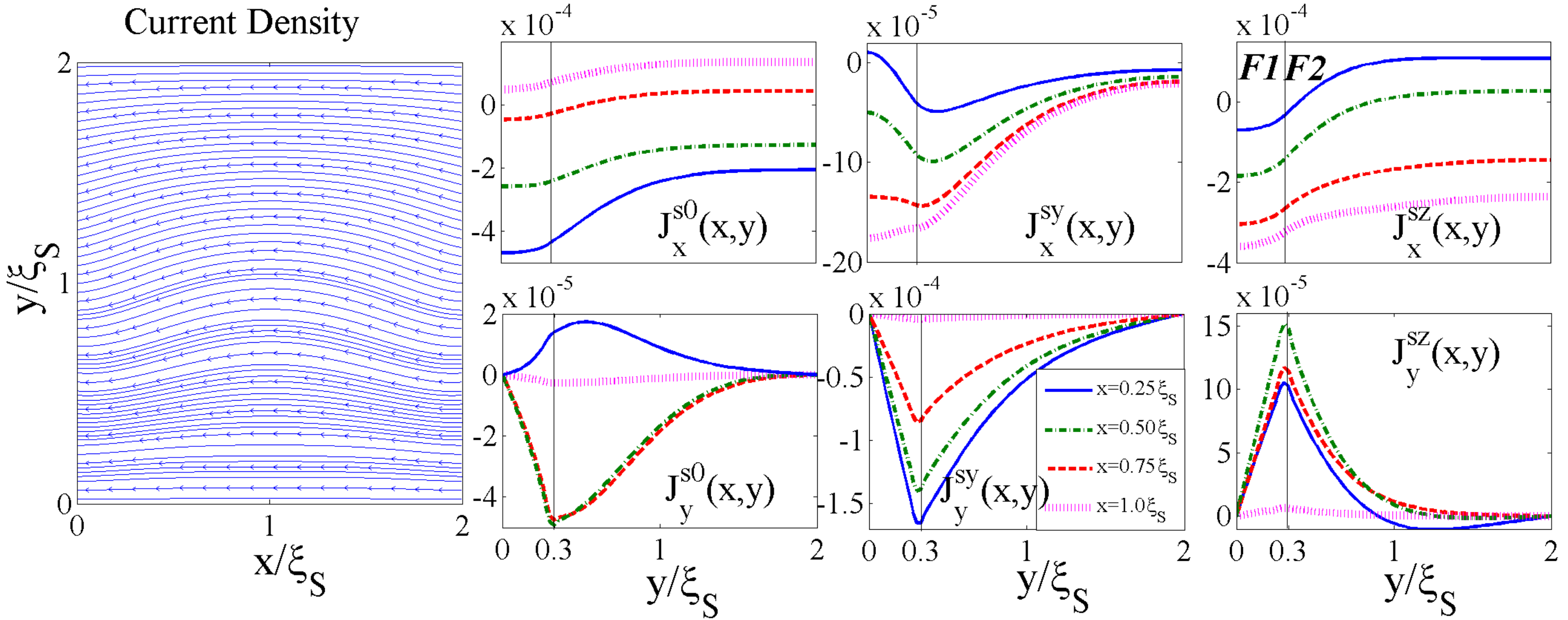}
\caption{\label{fig:jxy_37_Phi0pi_y} Spatial map of the critical
current density $\vec{J}(x,y)$ and its even- and odd-frequency
components. The configuration of the $S/F_1/F_2/S$
Josephson junction considered here is shown in Fig.
\ref{fig:model1}(a). The junction length $d_F$ is equal to
$2.0\xi_S$ and the width of each ferromagnetic strip labeled $F_1$
and $F_2$ are unequal, i.e., $W_{F1}$$=$$0.3\xi_S$ and
$W_{F2}$$=$$1.7\xi_S$. Vertical lines in the panels separate the two
ferromagnetic regions along the $y$ direction.}
\end{SCfigure*}
Our formalism allows for arbitrary magnetization orientation in each
ferromagnetic wire. However to most effectively realize the
long-range behavior of the decomposed charge supercurrent, we take
the magnetizations in the $F_1$ and $F_2$ wires to be orthogonal
\cite{bergeret_rmp}, with exchange fields $\vec{h}_1$$=$$(0,h^y,0)$,
and $\vec{h}_2$$=$$(0,0,h^z)$ or equivalently;
$\beta_1$$=$$\alpha_1$$=$$\pi/2$, and $\beta_2$$=$$0$. We focus here
on uniform $F_1/F_2$ bilayers (the least and simplest layered
structure) with $F_1$ and $F_2$ having unequal thicknesses as this
results in more pronounced singlet-triplet conversion
\cite{alidoust_sffn}. The magnitude of the exchange field in each
ferromagnet is typically set at $|\vec{h}_i|$$=$$10\Delta_0$, the
temperature $T$ corresponds to $T$$=$$0.05T_c$, and $\zeta$$=$$4$.
The energies are normalized by the superconducting gap at
zero temperature, $\Delta_0$, and the lengths by the superconducting
coherence length, $\xi_S$.

\begin{figure}[b!]
\includegraphics[width=8.7cm,height=8.0cm]{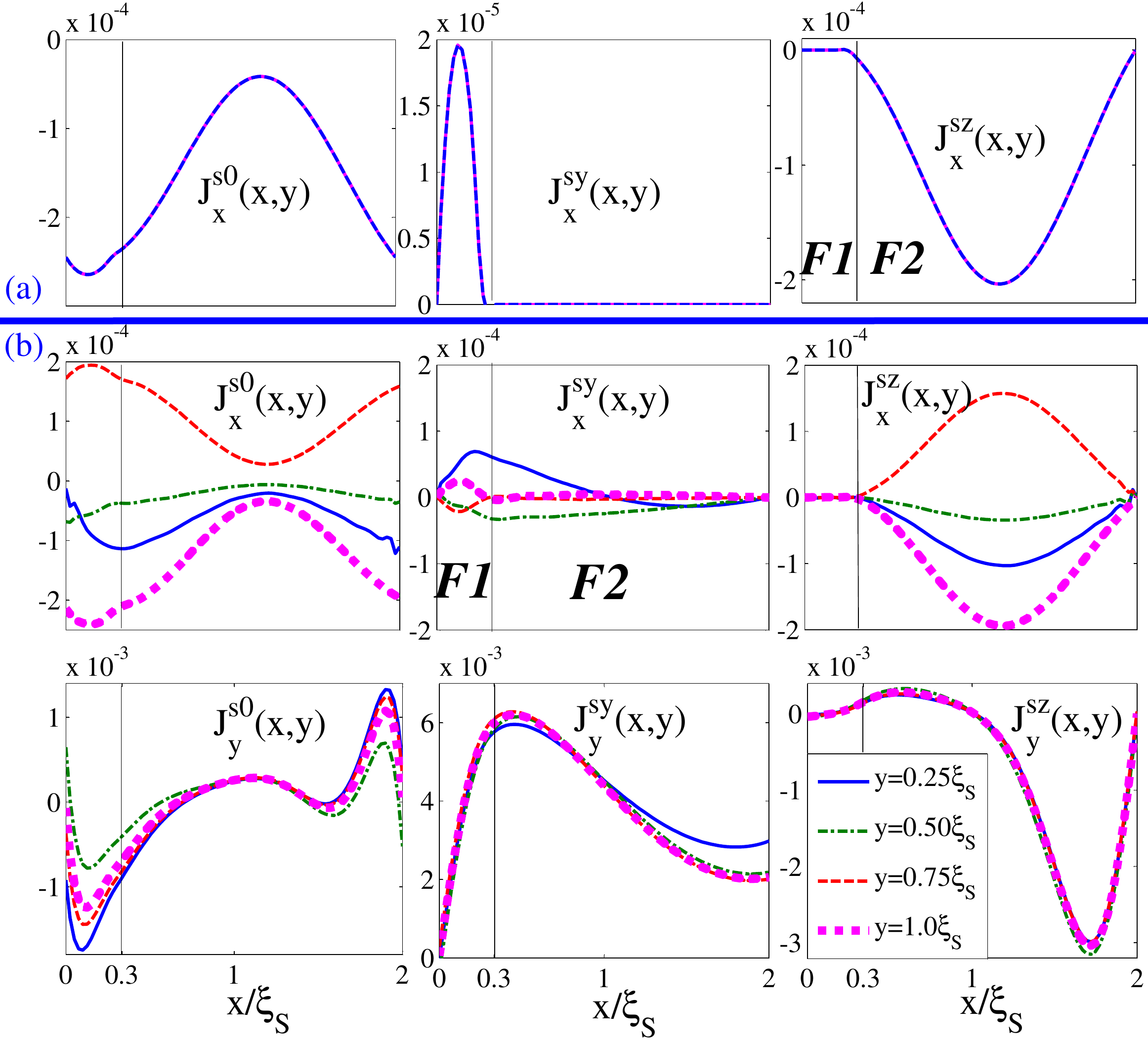}
\caption{\label{fig:Jxy_37_Phi3pi_x} Decomposed components of
critical supercurrent density in an $S/F_1/F_2/S$ junction depicted
in Fig.~\ref{fig:model1}(b). The two ferromagnetic strips have
unequal lengths: $d_{F1}$$=$$0.3\xi_S$, and $d_{F2}$$=$$1.7\xi_S$,
while their widths are equal: $W_{F1}$$=$$W_{F2}$$=$$2.0\xi_S$. In
part (a), there is no applied magnetic field, whereas in part (b)
the external magnetic field corresponds to a flux of
$\Phi$$=$$3\Phi_0$. The critical current components in both cases,
(a) and (b), are plotted as a function of $x$-position at four
different locations along the junction width: $y$$=$$0.25\xi_S$,
$0.50\xi_S$, $0.75\xi_S$, $1.0\xi_S$. }
\end{figure}

To have absolute comparisons, we constrain the rectangular $F$
strips to have equal total lengths and widths, i.e.,
$d_F$$=$$d_{F1}$$+$$d_{F2}$$=$$2.0\xi_S$ and
$W_F$$=$$W_{F1}$$+$$W_{F2}$$=$$2.0\xi_S$, respectively (see
Fig.~\ref{fig:model1}). The $S$ electrodes are given a macroscopic
phase difference of $|\varphi|$$=$$\pi/2$, which is the phase
difference corresponding to the critical supercurrent. In
Fig.~\ref{fig:jxy_37_Phi0pi_y} we show results corresponding to the
setup in Fig.~\ref{fig:model1}(a) with $W_{F1}$$=$$0.3\xi_S$,
$W_{F2}$$=$$1.7\xi_S$, and no applied magnetic field. First, the
spatial map of the total maximum charge supercurrent density shows
that the current in the vicinity of the $F_1/F_2$ junction has a
nonzero $y$-component \cite{crouzy} (as mentioned earlier,
$y/\xi_S$$=$$0.3$ coincides with the width of the $F_1$ nano-wire,
$W_{F1}$$=$$0.3\xi_S$). This result can be compared with
Fig.~\ref{fig:Jxy_37_Phi3pi_x}(a) for an  $S/F_1/F_2/S$ junction
with transverse supercurrent flow relative to the $F_1/F_2$
interface (e.g., Fig.~\ref{fig:model1}(b) with $\vec{H}$$=$$0$).
Figure~\ref{fig:jxy_37_Phi0pi_y} reveals that the induced
$y$-component is present over the entire junction width as exhibited
by the curved quasiparticle current trajectories. The spatial
behavior of the triplet and singlet contributions to the
supercurrent is also shown as a function of $y$ (along the junction
width) at four $x$ locations along the junction length. The top and
bottom set of panels corresponds to the current in the $x$ and $y$
directions respectively. The singlet and triplets components to the
supercurrent both play important roles in the supercurrent: The
amplitude of $J_{y}^{s0}$ is comparable to
$|J_{y}^{sy}$$+$$J_{y}^{sz}|$. These components of $J_y$, and hence
$J_y$ itself, vanish at $y$$=$$0$, $2.0\xi_S$, corresponding to the
vacuum borders. The vector plot of the supercurrent density in
Fig.~\ref{fig:jxy_37_Phi0pi_y} reveals that the current flow in the
middle of the junction ($x$$=$$\xi_S$) is directed entirely along
the $x$-direction throughout the junction width. The $x$-component
of the singlet contribution to the current, $J_{x}^{s0}$, becomes
vanishingly small when approaching this central region. On the
contrary, the triplet contributions $J_{x}^{sy}$ and $J_{x}^{sz}$
are maximal there, demonstrating optimal singlet-triplet conversion.
Another important aspect of this type of junction is seen in the
behavior of $J_{x}^{sy}$ and $J_{x}^{sz}$ as a function of $y$:
these triplet components generated in one $F$ region penetrate
deeply into the adjacent $F$ segment. Therefore, two important
phenomena arise: First, a singlet supercurrent flowing parallel to
the interface of the $F_1/F_2$ bilayers is converted into a triplet
supercurrent of both spin projections. Second, there is a deep
penetration of
odd-frequency triplet supercurrents laterally (along $y$) %khf
into the $F$ wires possessing orthogonal magnetizations relative to
the spin projection of the
triplet supercurrents. The signatures of long-ranged proximity
effects on the critical supercurrents flowing across similar
structures as Fig. \ref{fig:model1}(a) are studied in Ref.
\onlinecite{paral_sffs}. The planar structures with domain wall
patterns are also investigated in Ref. \onlinecite{paral_neel} where
the textured magnetizations generate long-ranged supercurrents
regardless of either `transverse' or `parallel' transports
\cite{bergeret_rmp}. We note that the results of Refs.
\onlinecite{paral_sffs} and \onlinecite{paral_neel} also fully
confirm the generic scenario introduced here.

Next we illustrate an $S/F_1/F_2/S$ junction depicted in
Fig.~\ref{fig:model1}(b). In Fig.~\ref{fig:Jxy_37_Phi3pi_x}(a), the
external magnetic flux is absent, while in (b) a magnetic flux
$\Phi$$=$$3\Phi_0$ is applied to the system. The geometric
dimensions correspond to $d_{F1}$$=$$0.3\xi_S$,
$d_{F2}$$=$$1.7\xi_S$, and the junction width, $W_F$$=$$2.0\xi_S$.
In the absence of an external magnetic field, the charge
supercurrent has no component along the $y$ direction
\cite{Cuevas_frh1,alidoust_nfrh,alidoust_nfrh1}, and is constant along $x$ (the
supercurrent flows transverse  to the $F$/$F$ interface
\cite{alidoust_sffn}). The relevant charge supercurrent density
components, $J_x^{s0}$, $J_x^{sy}$, and $J_x^{sz}$, thus only vary
spatially in the $x$ direction and their behaviors are substantially
different from their counterparts where the supercurrent flows
parallel to the $F_1/F_2$ interface [Fig. \ref{fig:model1}(a) setup]
shown in Fig.~\ref{fig:jxy_37_Phi0pi_y}. It is evident that
$J_x^{sy}$ disappears in $F_2$ while $J_x^{sz}$ is zero inside the
$F_1$ segment since we have the exchange fields
$\vec{h}_1$$=$$(0,h^y,0)$, and $\vec{h}_2$$=$$(0,0,h^z)$ acting
effectively as triplet spin filters. In other words, $J^{sy}$ is
generated in $F_1$, becomes localized there and then at the
$F_1/F_2$ interface, converts to the short-ranged $J^{sz}$ in $F_2$
(or vice versa). The
triplet supercurrent
generated in one $F$ is closely
linked to the local magnetization texture and therefore does not
penetrate into the other $F$ whose magnetization is orthogonal to
the spin direction \cite{alidoust_sffn,Trifunovic,crouzy}. Turning
now to Fig.~\ref{fig:Jxy_37_Phi3pi_x}(b), a magnetic field is
applied, corresponding to a magnetic flux of $\Phi$$=$$3\Phi_0$.
This induces proximity vortices that result in a nonuniform
supercurrent response varying in both the $x$ and $y$ directions
\cite{Cuevas_frh1,alidoust_nfrh,alidoust_nfrh1}. The corresponding supercurrent
constituents, $J^{s0}_{y}(x,y)$, $J^{sy}_{y}(x,y)$, and
$J^{sz}_{y}(x,y)$, are thus also nonzero along $y$. The amplitude of
$J^{sy}_{x}(x,y)$, is smaller than
the other components due in part to the small $F_1$ width,
$d_{F1}$$=$$0.3\xi_S$, consistent with
Fig.~\ref{fig:Jxy_37_Phi3pi_x}(a). As seen,
$J^{sz}_{x}(x,y)$, is drastically suppressed at the $F_1/F_2$
interface, similar to the case in Fig.~\ref{fig:Jxy_37_Phi3pi_x}(a).
Examining the singlet and triplet components to the supercurrent
flowing in the $y$-direction, we see that their contribution has
increased dramatically compared to propagation in the $x$ direction
(nearly an order of magnitude or more). The overall spatial behavior
of the even-frequency triplet components are different: The
component with orthogonal spin projection to the magnetization
direction $J^{sy}_{y}(x,y)$ penetrates now extensively into the
ferromagnetic regions compared with its $x$-component counterparts
and the component with parallel spin projection, $J^{sz}_{y}(x,y)$.
We note that the magnetization direction in $F_2$ is orthogonal to
the spin polarization of the $J^{sy}$ supercurrent component. By
considering these findings together with those of a $S/F_1/F_2/S$
configuration where the $F_1/F_2$ interface is orthogonal to the \fs
interface in Fig.~\ref{fig:jxy_37_Phi0pi_y}, and the results of
Refs. \onlinecite{paral_sffs} and \onlinecite{paral_neel}, one
concludes that: A singlet supercurrent flowing parallel to the
interface of a simple $F_1/F_2$ bilayer with noncollinear exchange
directions, can significantly convert to a long-ranged spin-triplet
supercurrent {\it regardless} of actual geometry.

\begin{figure}[t!]
%\raggedright
\includegraphics[width=8.5cm,height=2.8cm]{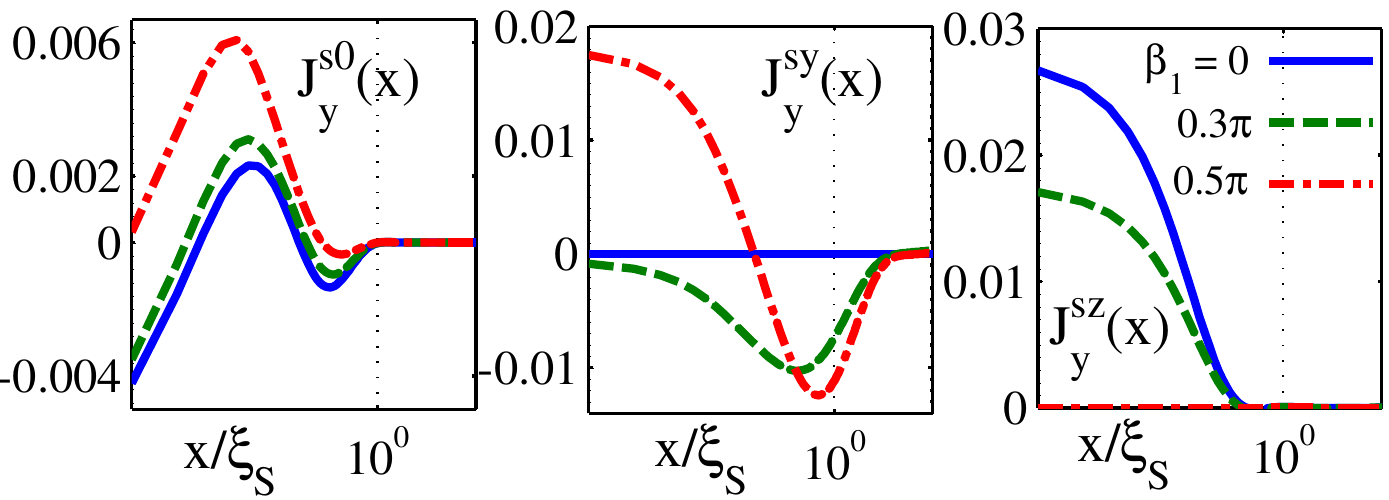}
\caption{\label{fig:sffn} Spatial behavior of decomposed diamagnetic
supercurrent as a function of lateral position inside the
$S/F_1/F_2/N$ junction depicted in Fig.~\ref{fig:model1}(c). Three
magnetization orientations of $F_1$ are shown: $\beta_1$$=$$0$,
$0.3\pi$, and $0.5\pi$ ($\alpha_1$$=$$\pi/2$), while $F_2$ remains
magnetized along $z$ ($\beta_2$$=$$0$). The thickness of the left
$F_1$ layer is fixed at $d_{F1}$$=$$0.1\xi_S$, while the thickness
of the right $F_2$ layer and $N$ metal are $d_{F2}$$=$$1.9\xi_S$ and
$d_N$$=$$2.5\xi_S$.}
\end{figure}

To further explore the generality of the phenomenon and %khf
provide an experimentally accessible platform to isolate it,
Fig.~\ref{fig:sffn} exhibits the diamagnetic supercurrent in a
$S/F_1/F_2/N$ structure [Fig.~\ref{fig:model1}(c)] with
$d_{F2}($$=$$1.9\xi_S)$$\gg$$ d_{F1}($$=$$0.1\xi_S)$ and
$d_N$$=$$2.5\xi_S$, at three magnetization orientations of $F_1$:
$\alpha_1$$=$$\pi/2$, $\beta_1$$=$$0$, $0.3\pi$, and $0.5\pi$, while
$\vec{h}_2$ points along the $z$ direction; $\beta_2$$=$$0$. As
seen,
$J_y^{sy}$  flowing along the $y$ direction, deeply penetrates
$F_2$ laterally.
Whereas, the singlet $J_y^{s0}$ and triplet $J_y^{sz}$ components
are short-ranged and drop drastically midway through the magnet
region. Hence, here is a clear and practical opportunity to generate
extensive long-range spin-triplet supercurrents by parallel
supercurrent flow relative to the interface of a uniformly
magnetized $F_1/F_2$ structure with inequivalent $F$ strips. The
$S/F_1/F_2/N$ structures proposed here are not only relatively
simple to fabricate and readily accessible to experimental
measurements \cite{robinson_sff}, but they also serve as an
efficient means to create samples for
probing the
previously discussed generic situations. In effect, the long-range
triplet components deeply penetrate laterally into the  thick %khf
$N$ region with minimal decay \cite{alidoust_sffn} while the singlet
and
short-range triplet supercurrent components nearly vanish by the
time they reach the $N$ layer. Therefore, this pure long-range
triplet current, which is odd in frequency, can be experimentally
isolated in very simple $S/F_1/F_2/N$ structures. Considering
todays' technological advancements, the experimental investigation
of the addressed phenomena in this Letter are readily accessible
\cite{robinson_sff}.

We would like to thank G. Sewell for valuable instructions in the
numerical parts of this work. We also appreciate
N.O. Birge for useful conversations and comments. K.H. is supported in part by ONR and
by a grant of supercomputer resources provided by the DOD HPCMP.


\begin{thebibliography}{000}


%--------------------------------------------------- triplet supercurrent ------------------------------------


\bibitem{zutic} I. Zutic, J. Fabian, and S.D. Sarma, \href{http://journals.aps.org/rmp/abstract/10.1103/RevModPhys.76.323}{\newblock \rmp \textbf{76}, 323 (2004).}


\bibitem{Kimura_prl} T. Kimura, Y. Otani, T. Sato, S. Takahashi, and S. Maekawa\href{http://journals.aps.org/prl/abstract/10.1103/PhysRevLett.98.156601}{\newblock \prl \textbf{98}, 156601
(2007)}.

\bibitem{golubov_rmp} A.G. Golubov, M.Y. Kupriyanov, and E. llichev, \href{http://journals.aps.org/rmp/abstract/10.1103/RevModPhys.76.411}{\newblock \rmp \textbf{76}, 411 (2004)}.

\bibitem{buzdin_rmp} A.I. Buzdin, \href{http://journals.aps.org/rmp/abstract/10.1103/RevModPhys.77.935}{\newblock \rmp \textbf{77}, 935 (2005)}.


\bibitem{Quay} C.H.L. Quay, D. Chevallier, C. Bena, and M. Aprili, \href{http://www.nature.com/nphys/journal/v9/n2/nphys2518/metrics}{\newblock Nat. Phys. \textbf{9}, 84 (2013)
};F. Hubler, M.J. Wolf, D. Beckmann, and H.V. Lohneysen,
\href{http://journals.aps.org/prl/abstract/10.1103/PhysRevLett.109.207001}{\newblock \prl \textbf{109}, 207001 (2012).
}.


\bibitem{bergeret_rmp} F.S. Bergeret, A.F. Volkov, and K.B. Efetov, \href{http://journals.aps.org/rmp/abstract/10.1103/RevModPhys.77.1321}{\newblock \rmp \textbf{77}, 1321
(2005)}.


\bibitem{ryaz0} R.S. Keizer, S.T.B. Goennenwein, T.M. Klapwijk, G. Miao, G. Xiao and A.
Gupta, \href{http://www.nature.com/nature/journal/v439/n7078/full/nature04499.html} {\newblock Nat. {\bf 439}, 825 (2006)} .

\bibitem{houzet} M. Houzet and A.I. Buzdin,
\href{http://link.aps.org/doi/10.1103/PhysRevB.76.060504} {\newblock \prb \textbf{76}, 060504(R)
(2007)}.

\bibitem{paral_neel} Y.V. Fominov,
A.F. Volkov, and K.B Efetov,
\href{http://journals.aps.org/prb/abstract/10.1103/PhysRevB.75.104509}
{\newblock Phys. Rev. B {\bf 75}, 104509 (2007)}.

\bibitem{norm} T.S. Khaire, M.A. Khasawneh, W.P. Pratt Jr., and N. Birge, \href{http://link.aps.org/doi/10.1103/PhysRevLett.104.137002} {\newblock \prl \textbf{104}, 137002
(2010)}.

\bibitem{robinson} J.W.A. Robinson, J.D.S. Witt, and M.G. Blamire, \href{http://www.sciencemag.org/content/329/5987/59
}{\newblock Science \textbf{329}, 5987 (2010)}; M. Alidoust, and J. Linder,
\href{http://link.aps.org/doi/10.1103/PhysRevB.82.224504} {\newblock \prb \textbf{82}, 224504 (2010)}.


\bibitem{paral_sffs} T.Y. Karminskaya,
M.Y. Kupriyanov, and A.A. Golubov,
\href{http://link.springer.com/article/10.1134/S0021364008100123}
{\newblock JETP Lett. {\bf 87}, 570 (2008)}; A.I. Buzdin, A.S.
Melnikov, and N.G. Pugach,
\href{http://journals.aps.org/prb/abstract/10.1103/PhysRevB.83.144515}
{\newblock Phys. Rev. B {\bf 83}, 144515 (2011)}.


\bibitem{Trifunovic} M. Houzet, A. I. Buzdin, \href{http://journals.aps.org/prb/abstract/10.1103/PhysRevB.74.214507}{\newblock
\prb  \textbf{74}, 214507 (2006)}; L. Trifunovic,
\href{http://journals.aps.org/prl/abstract/10.1103/PhysRevLett.107.047001}{\newblock
\prl \textbf{107}, 047001 (2011)}; L. Trifunovic, Z. Popovic, and Z.
Radovic,
\href{http://journals.aps.org/prb/abstract/10.1103/PhysRevB.84.064511}{\newblock
\prb  \textbf{84}, 064511 (2011) }.

\bibitem{alidoust_sffn} M. Alidoust, and K. Halterman, \href{http://mohammad-alidoust.blogspot.com}{\newblock A detailed paper including analytics and numerics: to be published (2015)}.

\bibitem{alidoust_nfrh} M. Alidoust, G. Sewell, and J. Linder,
\href{http://journals.aps.org/prl/abstract/10.1103/PhysRevLett.108.037001}{\newblock \prl \textbf{108}, 037001 (2012)}.
\bibitem{alidoust_nfrh1} M. Alidoust, and J. Linder,
\href{http://link.aps.org/doi/10.1103/PhysRevB.87.060503} {\newblock \prb {\bf 87}, 060503(R) (2013)}.

\bibitem{cite:usadel} K.D. Usadel,\href{http://prl.aps.org/abstract/PRL/v25/i8/p507_1}{\newblock \prl \textbf{25}, 507
(1970)}.

\bibitem{cite:zaitsev} M.Y. Kuprianov, Sov. Phys. JETP \textbf{67}, 1163 (1988).

\bibitem{crouzy} B. Crouzy, S. Tollis, D.A. Ivanov \href{http://journals.aps.org/prb/abstract/10.1103/PhysRevB.75.054503} {\newblock \prb \textbf{75}, 054503 (2007)}; ibid
\href{http://journals.aps.org/prb/abstract/10.1103/PhysRevB.76.134502}{\newblock
\textbf{76}, 134502 (2007)}.


\bibitem{Cuevas_frh1} J.C. Cuevas and F.S. Bergeret, \href{http://journals.aps.org/prl/abstract/10.1103/PhysRevLett.99.217002}{\newblock \prl \textbf{99},
217002 (2007)}.

\bibitem{robinson_sff} X.L. Wang, \etal, %A.D. Bernardo, N. Banerjee, A. Wells, F.S. Bergeret, M.G. Blamire, and J.W.A. Robinson
\href{http://journals.aps.org/prb/abstract/10.1103/PhysRevB.89.140508}{\newblock \prb \textbf{89}, 140508(R) (2014)}.


\end{thebibliography}
\end{document}